**Physical Laws Must be Invariant Over Quantum Systems**

Paul Merriam

*Abstract*
Decoherence may not solve all of the measurement problems of quantum mechanics. It is proposed that a solution to these problems may be to allow that superpositions describe physically real systems in the following sense. Each quantum system "carries" around a local spacetime *in whose terms* other quantum systems may take on nonlocal states. Each quantum system forms a physically valid coordinate frame. The laws of physics should be formulated to be invariant under the group of allowed transformations among such frames. A transformation of relatively superposed spatial coordinates that allows an electron system to preserve the de Broglie Relation in describing a double-slit laboratory system—in analogy to a Minkowskian Transformation—is proposed. In general, "quantum relativity" says $\hbar = 1$ is invariant under transformations among quantum reference frames. Some conjectures on how this impacts gravity and gauge invariance are made.

**Résumé**
Décohérence ne peut pas resoudre tous les problèmes de mesurage de la mécanique quantique. Une solution possible de ces problèmes est de permettre aux superpositions de décrire des systèmes physiques reéls de la manière suivante : chaque système quantique « emporte » un espace-temps locale, *dans les termes des lesquels* autre systèmes quantiques peuvent acquerir des êtats non- locaux. Chaque système quantique constitue une cadre de coordonnées valides. Les lois de physique doivent être énoncées comme invariants du group des transformation permis aux ces cadres. Une transformation des coordonnées spaciales relativement superposées est proposée , permettant à une systeme d'electrons de conserver la relation de *de Broglie* en décrivant une système de laboratoire de fantes d'Young- en analogie avec une transformation Minkovskienne. En général « la relativité quantique » affirme que de $\hbar =1$ est invariant pendant les transformation des cadres de reference quantiques. Des conjectures sont présentées concernant les possible répercussions de l'invariance de $\hbar =1$ sur la gravitation et l'invariance de jauge.

**Introduction**

How to comprehend superpositions and their relation to classical reality is at least a part of the measurement problem of quantum mechanics (Wheeler and Zurek 1983). The purpose of this paper is to make some observations about superpositions and the laws of physics, and the interpretation of quantum mechanics.

The basic idea is that if a laboratory describes an electron by de Broglie's relation, then the electron describes the laboratory by the same equation with the same constant. The laboratory would be too massive to generate the corresponding interference patterns in the electron's coordinates, necessitating Lorentz-like transformations based on relative energy.

Bub (1997) termed decoherence part of the "new orthodoxy" of quantum mechanics (Schlosshauer, 2004). The brief analysis in the next section argues for the conclusion that decoherence is a valid application of quantum mechanics that does not solve all foundational problems without further assumptions. Then the quantum invariance ideas are developed using similar notation, with foundational issues revisited. It is argued that if the quantum state of a physical system is *not* regarded as being relative to the reference system, then quantum mechanics is inconsistent or incomplete as a fundamental theory. Sections after that contain speculations on where quantum relativity leads. The paper does not inventory the resulting structures, the focus is on the 'quantum relativity' interpretation of quantum mechanics.

**Entangled Systems and Decoherence**

We start by modeling simple entangled systems, following Zurek 2003, Hughes 1989, and Jaunch 1968.

Suppose an "apparatus" system $A$ (this could be a Stern-Gerlach magnet, experimenter, etc.) describes some quantum mechanical system $S$ (e.g. an electron) in a Hilbert space $\mathbf{H}_{SA}$. Assume that $\mathbf{H}_{SA}$ is spanned by eigenvectors $|s_\uparrow\rangle$ and $|s_\downarrow\rangle$ of some observable represented by $\hat{S}$. Suppose that when $A$ observes (performs a measurement on) system $S$, it measures the value of $\hat{S}$. Then system $S$ will be found to be in state $|s_\uparrow\rangle$ or else in state $|s_\downarrow\rangle$. Now consider a third system, $E$ for "environment". Allow for the sake of argument that $E$ represents system $A$ in a Hilbert space $\mathbf{H}_{AE}$ (perhaps the measurement apparatus is a small molecule). $A$ may be observed to be in one of three orthogonal states: $|A_0\rangle$ representing $A$ before it observes system $S$, $|A_{S_\uparrow}\rangle$ representing $A$ after it has observed $S$ to be in state $|s_\uparrow\rangle$, and $|A_{S_\downarrow}\rangle$ representing $A$ after it has observed $S$ to be in state $|s_\downarrow\rangle$.[1]

---

[1] In discussions of Decoherence and quantum computing $A$ will often start in one of the measured states, e.g. $|A_0\rangle = |A_{S_\uparrow}\rangle$, and $A$ changes to state $|A_{S_\downarrow}\rangle$ only if state $|s_\downarrow\rangle$ is observed.

$S$ is allowed to start in an arbitrary superposition:

(1) $|S_0\rangle = c_1|s_\uparrow\rangle + c_2|s_\downarrow\rangle$.

The combined $SA$ system evolves (Schrödinger picture):

(2) $|S_0\rangle|A_0\rangle = (c_1|s_\uparrow\rangle + c_2|s_\downarrow\rangle)|A_0\rangle \rightarrow c_1|s_\uparrow\rangle|A_{S_\uparrow}\rangle + c_2|s_\downarrow\rangle|A_{S_\downarrow}\rangle = |\Psi_t\rangle$

where the $c_i$'s are a function of time and $|\Psi_t\rangle \in \boldsymbol{H}_{SE} \otimes \boldsymbol{H}_{AE}$ or for mixed states $|\Psi_t\rangle \subseteq \boldsymbol{H}_{SE} \otimes \boldsymbol{H}_{AE}$.[2]

Schlosshauer (2004) emphasizes two aspects of the measurement problem that Decoherence (reviewed in Zurek 2003) is often supposed to have resolved. First, it is not obvious how the classical world (with definite pointer positions) could emerge from superpositions. In Decoherence, $|\Psi_t\rangle$ continues to evolve unitarily, with the environment selecting relatively stable classical appearances. (It is sufficient for the program of Decoherence that quantum systems only appear to be classical, see Breuer 1996.) An alternative way of approaching these problems will be developed below.

Second is the "change-of-basis" problem. The change-of-basis problem is that it is possible to formally change the basis of the Hilbert space $\boldsymbol{H}_{SE} \otimes \boldsymbol{H}_{AE}$. The result is that $|\Psi_t\rangle$ can be written by $E$ as a superposition of states that have nothing to do with $A$ observing $S$ to be in state $|s_\uparrow\rangle$ or else state $|s_\downarrow\rangle$. For example, let

(3) $|s_\rightarrow\rangle = \frac{1}{\sqrt{2}}(|s_\uparrow\rangle + |s_\downarrow\rangle)$; $\quad |s_\leftarrow\rangle = \frac{1}{\sqrt{2}}(|s_\uparrow\rangle - |s_\downarrow\rangle)$

Then $|\Psi_t\rangle$ would be written as something like

(4) $|\Psi_t\rangle = c_3|s_\rightarrow\rangle|A_{ObservedState1}\rangle + c_4|s_\leftarrow\rangle|A_{ObservedState2}\rangle$

But obviously this doesn't make any sense; $A$ *didn't* measure the observable associated with the horizontal arrows. In Decoherence quantum systems evolve into "pointer bases" which correspond to those bases of $\boldsymbol{H}_{SE} \otimes \boldsymbol{H}_{AE}$ that appear to the system $A$ to be classical. This evolution satisfies the Schrödinger equation (in non-relativistic cases) but, crucially, only occurs after a decoherence time $\tau_D$ (Zurek 2003, p. 25). $\tau_D$ depends on various parameters like the environment's temperature and the number of ways $A$ can interact with the environment. Not to be lost sight of is the fact that it is not hard for $\tau_D$ to be huge compared to the Planck Time (See e.g. Mohanty and Webb 2003). For $0 < t < \tau_D$ Decoherence does not solve the change-of-basis problem.

---

[2] For the sake of consistency with what is explained below, I should say "the $c_i$'s are a function of $E$'s time."

**Intransitivity**

It is possible to make explicit another problem with (2). System $E$ is formally a "Wigner's-friend" kind of system, in that it stands outside of—and quantum mechanically describes—the apparatus system $A$. Thus at some times $t$

(5) the physical state of $SA$ according to $E$ is $|\Psi_t\rangle$

(6) the physical state of $SA$ according to $SA$ is $A_{S_\uparrow}$ or else $A_{S_\downarrow}$

and

(7) the states in (5) and (6) are not the same states.

Fortunately the parameter 'according to' changed from (5) to (6). Quantum Mechanics may be made consistent if the quantum state of a system (such as $SA$) is at least partially a function of which quantum system is describing it.[3] Before using this idea it is possible to look at (7) terms of more general conditions on physical theories. In fundamental theories of physics physical reality ought to form an equivalence class, namely

(8)  if $a$ is physical relative to $a$, then $a$ is physical relative to $a$
(9)  if $a$ is physical relative to $b$, then $b$ is physical relative to $a$
(10) if $a$ is physical relative to $b$, and $b$ is physical relative to $c$, then $a$ is physical relative to $c$

Define a relation x$Q$y = 'y is in a superposition relative to x', then the problem is that according to (7)

(11)   E$Q$A and A$Q$E do not imply E$Q$E

Therefore, unless quantum states are understood as relative to quantum system quantum mechanics is incomplete or inconsistent as a fundamental physical theory, since (11) does not describe an equivalence class.

**Quantum Relativity**

It is necessary to distinguish between two senses of space. In formulating quantum theories one often uses parameters $x$ and $t$. These parameters are specific to the ("local") quantum system. For the example above we may say $A$ describes $S$ as $|\Psi_S\rangle = |\Psi_S\rangle(x_A, t_A)$. In the quantum description is an operator $\hat{x}$. This operator is used to describe the physical relationship of system $S$ to system $A$. It can be used to compute expectation values, etc. $S$ may be non-local according to $x_A$, and $t_A$. But Non-locality in $A$'s terms does not imply $S$'s non-existence. The essential point of this paper is that since both systems physically exist they are both valid coordinate frames from which the laws of physics must hold. Quantum mechanics is as valid in $S$ as it is in $A$. Therefore $S$ will describe $A$ by a state vector $|\Psi_A\rangle = |\Psi_A\rangle(x_S, t_S)$. If $S$ is non-local in terms of $(x_A, t_A)$ then $A$ is non-local in terms of $(x_S, t_S)$.

In particular, $A$ describes $S$ as in (1). Therefore $S$ will describe $A$ as starting out in some corresponding superposition

---

[3] See [3], [11a], [15], and references therein. The notion that the quantum state of a system is relative to the quantum system describing it was called by Rovelli the *Relational State* analysis of quantum mechanics. Despite the names, the Relational State ideas are not the same as the "Relative State" analysis of Everett (1957), and (as will become implied) the "Relatively Objective Histories" analysis of Decoherence in Zurek (2003) pp. 44-46.

(12) $\quad |A_0\rangle = c_3|A_{S_\uparrow}\rangle + c_4|A_{S_\downarrow}\rangle$

where $\boldsymbol{H}_{SA}$ is isomorphic to $\boldsymbol{H}_{AS}$. When $A$ observes $S$ to be in some eigenstate $|s_i\rangle$ $S$ must also observe $A$ to be in some corresponding eigenstate $|A_i\rangle$ whence $|c_1|^2 = |c_3|^2$ etc.

This has consequences analogous to those of Special Relativity. For example, let $S$ be a free non-relativistic electron and $A$ be a double-slit experimental set up. From $A$'s point of view the probability amplitude $\Psi_{SA}$ of $S$ evolves according to the Schrödinger equation

(13) $\quad i\hbar \dfrac{\partial \Psi_{SA}}{\partial t} = -\dfrac{\hbar^2}{2m}\dfrac{\partial^2 \Psi_{SA}}{\partial x^2}$

The mass of $A$ is envisioned as being much larger than that of $S$. Therefore, the corresponding description of the probability amplitude $\Psi_{AS}$ of $A$ by $S$ would seem to violate *its* Schrödinger evolution

(14) $\quad i\hbar \dfrac{\partial \Psi_{AS}}{\partial t} = -\dfrac{\hbar^2}{2m}\dfrac{\partial^2 \Psi_{AS}}{\partial x^2}$

The de Broglie wavelength of $A$ would appear to be too small to allow an interference pattern, but that is illusory. Lengths do not have to have the same numerical values in both systems. (14) could be saved if two quantum observations are $\Delta x$ apart in $A$ but $\Delta x'$ apart in $S$ such that

(15) $\quad \Delta x' \approx \Delta x \sqrt{\dfrac{m_S}{m_A}}$

where $m_A$ is the mass of system $A$ and $m_S$ is the mass of system $S$ (up to a constant) obtains. The coordinates of $S$ are dilated (by a factor $>1$) in terms of the coordinates of $A$.[4] This allows $S$ to maintain the de Broglie relation

(16) $\quad m'v'\lambda' = \hbar$

in describing $A$. The numerical value of $\lambda'$ is small, but the length $\Delta\lambda'$ is magnified by $\sqrt{\dfrac{m_A}{m_S}}$ in terms of $A$'s spatial coordinates. There is a group of such transformations. This is the group of transformations among $(x_Q, t_Q)$ for arbitrary allowed quantum systems $Q$. Form-invariance of the equations of physics under the group action is possible if one "form" is any representative of the equivalence class of expressions $f_i, f_i \otimes f_j, f_i \otimes f_j \otimes f_k, ...$ that give the same amplitude at a given local amplitude function $f$.

Writing $\hbar^{-1}$ in analogy to $c$, the dimension [Joules$^{-1}$] is analogous to [meters], and (allowing "$J_i$" for energy) taking into account $\Delta\lambda = \dfrac{1}{J_2} - \dfrac{1}{J_1} = \dfrac{J_1 - J_2}{J_1 J_2}$ the invariant quantum interval $\upsilon$ is

$$\text{(16b)} \quad \upsilon = \left[ (\Delta t)^2 + \left( \frac{\Delta \lambda}{\lambda_1 \lambda_2} \right)^2 \right]^{\frac{1}{2}}$$

---

[4] This makes intuitive sense (especially in the position basis) as a smaller system would have a larger associated de Broglie wave which would be less fine of an "instrument" to prepare and probe states of other systems.

It has positive signature and reduces to the Minkowskian relativistic interval $\tau$ via $[J]$ = $[m^2 \cdot kg \cdot s^{-2}]$.

In analogy to

$$\text{(16c)} \quad \gamma = \frac{1}{\left(1 - \frac{v^2}{c^2}\right)^{\frac{1}{2}}}$$

it is easy to geometrically derive

$$\text{(16d)} \quad \delta = \frac{1}{\left(1 - (E_q t / h)^2\right)^{\frac{1}{2}}}$$

with $E_q$ the relative quantum energy, and should be done in the context of quantum systems. In general the equation $\hbar = 1$ should be invariant over relatively quantum systems and interpreted the same way as the invariance of $c = 1$ over relatively moving systems.[5]

---

[5] Imposing $c = 1$ can be seen as a rescaling of pre-relativistic measures of durations, and applies to Galilean or Leibnizian relational conceptions of time. Its relation to Newton-Barrow absolute becoming is less clear.

**Measurement Problems in Quantum Relativity (Part 1)**

The first measurement problem was that of how classical appearances emerge. In "quantum relativity" there is on the one hand each systems' quantum description (in that system's local variables of space $x$ and time $t$) of other quantum systems. During two systems' quantum interaction (performing a measurement on each other) they are considered the same quantum system and therefore share a single space-time frame. (I believe this is similar in Rovelli [2004]). This opens the question of the ontological coordination of relatively superposed systems, i.e. systems that are not sharing a space-time frame. See the "Gravity" section below.

The second measurement problem was the change-of-basis problem. In the quantum relativity approach the projection postulate is not necessarily associated with a quantum-classical divide, but rather it is the formal representation of two quantum systems in the process of interacting, as described by one system or the other. Before (according to E's time) $E$ actually observes the system described by $|\Psi_t\rangle$ then the system is still in a superposition (relative to $E$), in which case it doesn't make any difference what basis $E$ chooses for $\boldsymbol{H}_{SE} \otimes \boldsymbol{H}_{AE}$. At the time $t_E$ that $E$ observes $SA$ (via the projection postulate) the relational requirement of inter-systemic observational agreement gives (using the intermediate notation of (4))

$$(17) \quad |S_0\rangle |A_0\rangle = (c_1 |s_\uparrow\rangle + c_2 |s_\downarrow\rangle) |A_0\rangle \xrightarrow{Schrodinger Evolution} c_3 |s_\rightarrow\rangle |A_{S_\uparrow}\rangle + c_4 |s_\leftarrow\rangle |A_{S_\downarrow}\rangle =$$

$$|\Psi_t\rangle \xrightarrow{\text{Pr} ojection Postulate} |s_\uparrow\rangle |A_{S_\uparrow}\rangle \text{ or else } |s_\downarrow\rangle |A_{S_\downarrow}\rangle.$$

after observations have definitely been made, with $c_i$ a function of E's time. Any observables system $S$ might choose to measure are constrained by $A$.

When $x$ is a parameter, as in $L(x, t)$ with $L$ some Lagrangian, it is (trivially) a coordinate of the local spacetime. When $\hat{x}$ is an operator describing another quantum system it is constrained. The freedom to choose which observable to measure is mathematically constrained by the future interaction between the systems. This is satisfactory so far as ontology goes; the analogy with Special Relativity will make this clear. Suppose some (non-quantum) system $X$, perhaps a proton, collides with some (non-quantum) system $Y$, say a neutron. Their relative velocity *before* the collision can only take on certain values (depending on the equations of motion) to be consistent with the energy and angle of the collision when they eventually do collide. Before the collision $X$ and $Y$ are in a state of relative motion. Before the observation $E$ and $A$ are in a state of relative superposition.

"Before" (as defined in system $E$) observation of the system $AS$, $AS$ is in a state of superposition relative to $E$, namely $|\Psi_t\rangle_{AS}$, where the "$t$" is a value in $E$'s spacetime. This is the "$|\Psi_t\rangle$" at the end of (2). We don't usually notice a lack of choice of observable when $A$ is macroscopic probably because of the huge difference in information between $A$ and $S$. On the other hand, when $A$ is taken to be the size of a nucleus (say), $|\Psi_t\rangle$ is taken to represent the usual entangled state. "Before" (as defined in system $A$) the interaction between $A$ and $E$, then $A$ describes $E$ as being in a correlated superposition = $|\Psi_t\rangle_{\text{environment}} = c_5 |s_\uparrow\rangle |E_{S_\uparrow}\rangle + c_6 |s_\downarrow\rangle |E_{S_\downarrow}\rangle$, where "$t$" is now a value in $A$'s spacetime.

The superposition $|\Psi_t\rangle_{\text{environment}}$ results (in this case) because before (in *A*'s time) *A* interacts with *E and* before (in *A*'s time) *A* interacts with *S* then *S* is as in (1) and unitary evolution (in terms of *A*'s $x$ and $t$) leads to *S*'s entanglement with *E*. It does not matter that *E* and *A* give each other different times of correlation (between *S* and *A* on the one hand and *S* and *E* on the other) because they have their own quantum parameters $x_E$, $t_E$ and respectively $x_A$, $t_A$ to begin with. (15) is a possibility for one transformation.

The third measurement problem was expressed in the example where the physical state of *SA* is $|\Psi_t\rangle$ according to *E* but $A_{S_\uparrow}$ or else $A_{S_\downarrow}$ according to *SA*. In the view of this paper, they can both be right, subject to the constraint of correlated observables and their relationally-relative values. An analogue in Special Relativity is constrained but different values for a length, depending on (relative) velocity.

**Gravity**

"... assume the complete physical equivalence of a gravitational field and a corresponding acceleration of the reference system." (Einstein, 1907) It makes sense to assume this for the "local" reference system. The local reference system is the local *quantum* system. The correct theory of quantum gravity starts with the theory of quantum observations above and applies the equivalence principle.

There are 5 parameters to a quantum measurement—amplitude, phase, and location. The quantum equivalence principle says that for each of these a gravitational field is physically indistinguishable from the corresponding acceleration of the reference system. General Relativity suggests there is a quantum Einstein's Equation

(18) $R_{AB} = 0$

on a 5-manifold, analogous to the classical Einstein's Equation on a 4-manifold, where distance looks like

(19) $\Delta v^2 = \Delta A^2 + \Delta B^2 + \Delta C^2 + \Delta D^2 + \Delta E^2$

with a boundary condition, for locally flat geometry. This implies a relation to the Standard Model, e.g. an electron cannot locally tell the difference between geometric acceleration and radiating a photon.

Meters and seconds can be measured in radians appropriately scaled. Relatively superposed systems carry around separate spacetimes, as mentioned earlier, in the sense of spacetime 4-hyperplanes of the 5-manifold.

**Measurement Problem (Part 2)**

Whether and why is a classical apparatus is needed for a quantum measurement? The value of an observable is classically measured of a thing if there is zero (relative) distance along the quantum dimension of the manifold between the thing and the reference quantum system, i.e. $R^2=1$. Otherwise the systems have moved off the (same) classical 4-manifold to the boundaries of a quantum 5-volume (assuming they interact again). Two systems with a *any* constant relative quantum separation satisfy an Einstein-type equation as a level surface of the 5-metric (and obviously there are other level surfaces). Nonlocality occurs ("trivially") in classical spacetime because the quantum dimension is locally perpendicular to classical spacetime—the electron is not moving *in* the local quantum system's classical spacetime. The 5-manifold is isomorphic to (Minkowski)×(quantum dimension).

Schrödinger's cat is in a relatively superposed state according to the laboratory spacetime if and only if the laboratory is in a corresponding superposition relative to the cat's spacetime. What state is Schrödinger's cat in? Relative to what system?

Do we need a classical apparatus to register a quantum interaction? The apparatus is "classical", i.e. has relative q-distance of 0 on the manifold, only relative to some systems. The question can be understood the same way as the question "do we need a non-relativisticly moving apparatus to register a relativistic interaction?"

**Gauges and Matrices**

Returning to a more 4-dimensional perspective for a moment, physical laws should be invariant in all gauges together. Here consider that a global (meaning spanning the reference quantum system's spacetime) absolute phase can be added to the amplitudes of quantum field operators without affecting any expectation values. This is implied by (15) in the form $\frac{\Delta x'}{\Delta t'} \propto \int_{5-space} \frac{\Delta x}{\Delta t}$ as it is a relationship between the other system's parameters $x'$ and $t'$, and not the parameters' "absolute" values that is transformed. The arbitrariness of absolute phase, in other words, results because the relation between $x'$ and $t'$ on the one hand, and $x$ and $t$ on the other, is invariant if, e.g. $x' = -i\,x$ and $t' = t$ or if we say $x' = x$ and $t' = it$.

Let $H_{Q_i Q_j}$ be the hilbert space on which quantum system $Q_j$ describes quantum system $Q_i$. Assume $H_{Q_i Q_j}$ is spanned by orthogonal basis vectors labeled by spacetime events as given by $Q_j$. These events are the most recent record $Q_j$ has of previous quantum interactions. Suppose now $Q_i$ describes quantum system $Q_h$ in a Hilbert space $H_{Q_h Q_i}$. Each basis vector $|\mathbf{x}\rangle$ should be converted to the integral cited in (15) above over the Hilbert space $H_{Q_i Q_j}$. Let $\{H\}_{Q_n Q_{n-1} \cdots Q_1}$ be a hilbert space modeling the relative superpositions of $n$ superposed systems. It is constructible by starting with the hilbert space $H_{Q_1}$ of quantum system $Q_1$. Moving to $H_{Q_2 Q_1}$, each position basis vector $|\mathbf{x}\rangle$ in the rigged $H_{Q_1}$ must be itself turned into a rigged hilbert space $|\mathbf{x}\rangle \to H_{Q_1}(|\mathbf{x}\rangle)$.

The correct interpretation of the axiom of quantum mechanics

(20) $\int |\Psi|^2 dx = 1$

where $dx$ ranges over all of space is that if *everywhere* were suddenly tested, the electron would *definitely* be found. The distinction between that interpretation and the (incorrect) interpretation of (20) as the $\Psi$-function "being everywhere" is that in the former "everywhere" means in terms of the local quantum system's spacetime and is therefore is not creating more spacetime "events" with the basis "Hilbertization" of basis vectors in the sense of (20) over the hilbert space $H_{Q_3 Q_2}$. The second interpretation of (20) would require, from the perspective of $Q_1$, the (incorrect)

$$(21) \int \left|\Psi_{Q_2 Q_1}\right|^2 dx_{Q_2 Q_1} = \int \left|\int \Psi_{Q_3 Q_2} dx_{Q_3 Q_2}\right|^2 dx_{Q_2 Q_1} = \int \left|\int ... \int \Psi_{Q_n Q_{n-1}} dx_{Q_n Q_{n-1}} ... dx_{Q_3 Q_2}\right|^2 dx_{Q_2 Q_1}$$

for the case of $n$ systems. Obviously, if general relativistic spacetime is a background upon which all quantum events happen with normalized probability, then (21) gives the QFT overestimate for the gravity of a field. If (21) takes place on a nested set of hilbert spaces $\{H\}_{Q_n Q_{n-1} ... Q_1}$ then, to begin with, the Schrödinger equations

$$(22) \quad i\hbar \frac{\partial \Psi_{Q_n Q_{n-1}}}{\partial t} = -\frac{\hbar^2}{2m} \frac{\partial^2 \Psi_{Q_n Q_{n-1}}}{\partial x^2}$$

imply an equation

$$(23) \quad \sum_{j=2}^{n} (i\hbar)^j \frac{\partial \Psi_{Q_j Q_{j-1}}}{\partial t} = \sum_{j=2}^{n} \left(-\frac{\hbar^2}{2m}\right)^j \frac{\partial \Psi_{Q_j Q_{j-1}}}{\partial x_{Q_{j-1}}}$$ which constrains the evolution between the

relative amplitude $Z_1$ and $Z_n$ between systems $Q_1$ and $Q_n$

$$(24) \quad \frac{\partial Z_{Q_n Q_1}}{\partial t} = k \frac{\partial^2 Z_{Q_n Q_1}}{\partial (x_{Q_1})^2}$$

$k$ some constant.

    A quantum system is locally able to give the evolution of relativistic quantum systems in terms of a spacetime hyperplane that, localized to quantum system, appears as Galieanian time.[6] I interpret the arguments of [Myrvold 2003] to justify this conclusion for relativistic quantum field theories.

    From the path-integral approach of integrating through all possible classical paths in spacetime, ghosts[7] are the expression of integrating over paths that form a closed loop, not in a single spacetime but among what are actually a number of quantum systems $Q_n$, ... $Q_1$. (The theory of measurement.)

    Quantum tunneling is like when an object in 4-d spacetime goes over another 4-object, through other spacetime hyperplanes on the 5-manifold. Quantizing a theory is the processes of moving (translating) the classical field into coordinates of a relatively superposed physical system. Corrections below relative action $\hbar$ (above relative information $I(\hbar^{-1})$) are like trying to specify a relativistic (classical) state perturbativly using terms above relative classical speed $c$. Gauge invariance results from the invariance of laws over different quantum systems.

**Conclusions**

---

[6] Time, defined as correlations among observations (such as propertime), is ontologically sufficient if the apparent direction of time is all there is to explain about time. But the question of the apparent direction of time is irrelevant to the deeper question of the apparent *uniqueness* of the present moment, and (the author argues in a related paper) thermodynamic or statistical correlations explain at most the former.

[7] Ghosts are not physically real, see e.g. Tony Zee's "Quantum Field Theory in a Nutshell" pp. 354-356 and Michio Kaku's "Quantum Field Theory" p. 304 and p.412.

One of the postulates of special relativity is that physical laws should be form-invariant under allowed changes of coordinate system. Ultimately this idea is not limited to relatively inertial systems. The coordinate systems we are interested in, so far as physical laws are concerned, are the coordinate systems justified by physically realizable states.[8] When *A* describes *S* as being in a superposition then *S* can *physically* have unsharp values of observables defined in *A*. That does not mean *S* has ceased to exist (nor to be physical). Therefore, laws valid in *A* should be form-invariant when translated to the reference frame of the superposition (1).

Philosophically, ontological reality must form an equivalence class. This holds whether reality is ultimately material or something else, and reduces to the case of whether one thinks of objects as superposed or not. The laws of physics must be made invariant under translations among relative superpositions because superpositions describe ontologically real coordinate systems. De Broglie's relation is not valid in the laboratory system unless it is valid in the electron system. A quantum relativity is required by (11).

Relatively superposed systems should leave the equation $\hbar = 1$ invariant. Equations should be derived to be invariant under the transformations among quantum coordinate frames. The tenant that the laws of physics should be coordinate-system-independent holds for quantum coordinate systems. Asking about the quantum state of Schrödinger's Cat is like asking about the relativistic velocity of a cat.

Application of the equivalence principle to the 5 parameters of a quantum measurement implies a relation between the Standard Model and the 5-manifold. In terms of the 5-manifold, a relatively superposed system is a subset of a different spacetime 4-hypersurface. Observations occur according to the evolution of the geodesics that locally leave the equations $c = 1$, $\hbar = 1$, etc. invariant.

Uncertainty relations are taken by some authors to be the essence of quantum strangeness. Here they are just standard deviations of measurements that are defined only to within a certain region of the 5-manifold, and do not have any non-classical strangeness.

A general mathematical investigation is possible, regardless of the contemporary state of physics, that would give the most general law *L* among objects $u_i \in U$ in the universe of a formal theory *T* such that $u_i$ are taken to ontologically exist, as expressed in *T*. The relations *L(M)* are then the most general that leave *L*, *T*, and *M* well-defined when expressed in the coordinates of each $u_i$.

---

[8] Classical general relativity is Machian when the gravitational field's own gravitating energy is taken into account, leading to e.g. frame dragging, see Rovelli [2004].

**Acknowledgements**

The author would like to acknowledge his indebtedness to R. Eric, D. Finkelstein, A. Frebert, B. Rosenblum, C. Rovelli, M. Schlosshauer, J. Schwartz, and an anonymous referee.

**References:**

1. R. Arnowitt, S. Deser, and C. Misner, Phys. Rev. 113, 745-750, 1959.

2. A. Aspect, J. Dalibard, and G. Roger, Phys. Rev. Lett. **49**, 1804 (1982) [Issue 25-20 December].

3. G. Bene, *Physica* **A** 242 (1992): 529-560.

4. D. Bohm, B. J. Hiley, (1993) *The Unidivided Universe*, Routledge, London and New York.

5. T. Breuer, Synthese **107**, 1 (1996).

6. A. Einstein (1905), (Methuen and Company, Ltd. of London, 1923).

7. A. Einstein (1907), translated "On the relativity principle and the conclusions drawn from it," in *The collected papers of Albert Einstein. Vol. 2 : The Swiss years: writings, 1900–1909* (Princeton University Press, Princeton, NJ, 1989), Anna Beck translator.

8. H. Everett III[rd], Rev. Mod. Phys. **29**, 454 (1957).

9. R. I. G. Hughes, *The Structure and Interpretation of Quantum Mechanics* (Harvard University Press, Cambridge, MA, 1989). Especially chapter 9, section 6.

10. J. M. Jaunch, *Foundations of Quantum Mechanics* (Addison Wesley, Reading, MA, 1968). Especially chapter 6, section 9.

11. R. Kjellander (1981) A geometrical definition of spinnors from 'orientations' in three-dimensional space leading to a linear spinor visualization. J. Phys. A: Math. Gen. **14** 1863-1885.

11. F. Markopoulou, L. Smolin. Online as http://arxiv.org/abs/gr-qc/9702025.

12. P. Merriam, e-journal Metaphysical Review, June 23, 1997.

13. P. Mohanty and R. A. Webb, Phys. Rev. Lett. **91**, 066604 (8 August 2003).

14. J. M. C. Montanus, Springer: ISSN: 0015-9018 (Paper) 1572-9516 (Online), Volume 31, Number 9, Date: September 2001, Pages: 1357 – 1400.

15. W. C. Myrvold, *British Journal for the Philosophy of Science*, September 2003, vol. 54, no. 3, pp. 475-500(26). Oxford University Press.

16. B. Rosenblum and F. Kuttner, online as quant-ph/0011086.

17. C. Rovelli, Cambridge University Press., Cambridge, MA, 2004. Especially section 5.6. See also online at http://plato.stanford.edu/entries/qm-relational/.

18. M. Schlosshauer, Rev. Mod. Phys. **76**, 4, online as quant-ph/0312059.

19. L. Smolin, arXiv:hep-th/0408048 v2 8 Aug 2004.

20. The Royal Swedish Academy of Sciences, "Advanced information on the Nobel Prize in Physics 2005", p.6, Oct. 4[th] 2005. This is online at http://nobelprize.org/.

21. W. Weiss and C. D'Mello, http://www.math.toronto.edu/~weiss/model_theory.html

22. P. Wesson, gr-qc/0205117

23. J. A. Wheeler and W. H. Zurek, eds., *Quantum Theory and Measurement* (Princeton Univ. Press, Princeton, NY, 1983).

24. R. D. Sorkin, A Kaluza-Klein Monopole, IAS 1983.

25. W. H. Zurek, Rev. Mod. Phys. **75**, 715 (2003). Online at arXiv:quant-ph/0105127 v3 19 Jun 2003.